\documentclass[aip,amsmath,amssymb,reprint]{revtex4-1}

\usepackage{graphicx}
\usepackage{dcolumn}
\usepackage{bm}
\usepackage[normalem]{ulem}
\usepackage[caption=false]{subfig}
\usepackage[utf8]{inputenc}
\usepackage[T1]{fontenc}
\usepackage{mathptmx}
\usepackage{etoolbox}
\usepackage{xcolor}
\definecolor{linkblue}{RGB}{0,70,150}
\definecolor{citered}{RGB}{170,40,40}

\usepackage[
    colorlinks=true,
    linkcolor=linkblue,  
    citecolor=citered,   
    urlcolor=linkblue
]{hyperref}

\newcommand{\D}{\mathrm{d}}
\newcommand{\e}{\mathrm{e}}

\newcommand{\be}{\begin{equation}}
\newcommand{\ee}{\end{equation}}
\newcommand{\bea}{\begin{align}}
\newcommand{\eea}{\end{align}}
\newcommand{\ba} {\begin{align} }
\newcommand{\ea} {\end{align} }

\newcommand{\kbt}{k_{\mathrm{B}}T}

\makeatletter
\def\@email#1#2{%
 \endgroup
 \patchcmd{\titleblock@produce}
  {\frontmatter@RRAPformat}
  {\frontmatter@RRAPformat{\produce@RRAP{*#1\href{mailto:#2}{#2}}}\frontmatter@RRAPformat}
  {}{}
}%
\makeatother
\begin{document}

\preprint{AIP/123-QED}

\title[]{Swimming versus Screening: How Activity Reshapes the Ionic Cloud of a Charged Particle}
\author{Tanmoy Chakraborty}
\email{tanmoy.statphys@gmail.com}
\affiliation{Department of Physics, Technion - Israel Institute of Technology, Haifa 320003, Israel}

\author{Tomer Markovich}
\email{tmarkovich@tauex.tau.ac.il}
\affiliation{School of Mechanical Engineering, Tel Aviv University, Tel Aviv 69978, Israel}

\author{Ram M. Adar}
\email{radar@technion.ac.il}
\affiliation{Department of Physics, Technion - Israel Institute of Technology, Haifa 320003, Israel}

\begin{abstract}
A major question in active matter is how interactions between particles are modified by activity. Here, we address this question for Coulomb interactions by considering a charged swimmer immersed in an electrolyte. A static charge is screened by an ionic cloud that decays over the Debye length $\lambda_D$. We show that the persistent motion of the swimmer can qualitatively modify the steady-state ionic charge-density profile. At low activity, the swimmer appears effectively quasistatic to the ions, and conventional Debye-–Hückel screening is recovered. At higher activity, the ballistic-to-diffusive dynamics of the swimmer are reflected in distinct spatial regimes. A power-law decay is followed by a stretched-exponential decay at intermediate distances. At sufficiently large distances, exponential screening is restored, but with an increased screening length, $\lambda>\lambda_D$, revealing activity-induced underscreening. Thus, active motion modifies both the spatial form and the range of electrostatic screening. Our results motivate the study of active electrolytes composed of positively and negatively charged swimmers.

\end{abstract}

\maketitle

\section{\label{sec:intro}Introduction}

Active matter is composed of particles that convert energy into persistent motion \cite{Ramaswamy_2010, Marchetti_RMP_2013, Bechinger_RMP_2016}. At times shorter than the persistence time, their motion is approximately ballistic, whereas at longer times it becomes diffusive. One of the central questions in the field is how this persistent motion modifies the response to external potentials and interparticle interactions. Even when the large-scale behavior is described only by scalar fields, without macroscopic orientational order, standard interactions can produce organizations with no equilibrium counterpart. Examples include motility-induced phase separation, rectification, and long-ranged correlations induced by boundaries and inclusions \cite{MIPS_Cates_2015, Reichhardt_ratchet_2017, metzger2026revisiting, metzger2026revisitingE, Granek_RMP_2024}.

For active particles immersed in a fluid, momentum is conserved by the combined particle--fluid suspensions. Force- and torque-free self-propelled particles in such systems are commonly referred to as \emph{swimmers} \cite{Lauga_2009, Elgeti_2015}. They include motile microorganisms and synthetic catalytic or phoretic colloids. Below, we use the term \emph{charged swimmer} for a swimmer that carries a net electric charge.

Interactions in scalar active matter can also be mediated by fields produced by the particles. In quorum-sensing models, a locally produced signal controls particle motility and couples the dynamics to the local density or chemical environment \cite{TailleurCates_PRL_2008, Lefranc_PRX_2025}. In phoretic systems, particles produce or consume a solute and move in response to its gradients. The resulting interactions are long-ranged and, in mixtures, generally nonreciprocal \cite{IllienGolestanianSen_CSR_2017, IllienGolestanian_2024,NasouriGolestanian_PRL_2020}. In both cases, particle motion modifies a mediating field, which in turn acts back on the particles.

An especially appealing mediating field is that of ions. Ions have been considered in active systems mainly in the context of self-phoretic swimmers, where asymmetric production or absorption of ions generates propulsion through coupled ionic, electrostatic, and hydrodynamic fields \cite{AsmolovVinogradova_POF_2023, ShresthaOlvera_PRE_2024}. In these systems, the ionic fields are part of the propulsion mechanism. Frydel and Podgornik considered an active electrolyte in which the mobile ions themselves are active and accumulate near a fixed charged wall, leading to dynamic adsorption and overscreening \cite{frydel2018}. In their problem, activity belongs to the screening charges and the source is fixed. Here, the roles are reversed; the source charge is active, while the screening ions remain passive. This reversal introduces a different mechanism: because the ionic cloud relaxes only over a finite time, the temporal structure of the swimmer motion can be imprinted on the spatial form and range of screening. 
 To our knowledge, the self-consistent screening of a charged swimmer has not been considered. 

In a dilute equilibrium electrolyte, a static point charge is surrounded by an ionic cloud that screens its Coulomb field \cite{verwey1948theory,andelman1995electrostatic,levin2002electrostatic,safran2003statistical,israelachvili2011intermolecular,markovich2021charged}. Debye--H\"uckel (DH) theory \cite{debye1923theorie1,debye1923theorie2} predicts a spherically symmetric cloud with Debye length $\lambda_D\equiv\kappa_D^{-1}=\sqrt{\epsilon k_BT/(2e^2n_b)}$, where $\epsilon$ is the solvent dielectric constant, $\kbt$ is the thermal energy, $-e<0$ is the electron charge, and $n_b$ is the bulk concentration of a monovalent salt. The electrostatic potential induced by a point charge $q$ placed at the origin is then
\begin{equation}\label{debye_huckel_potential}
\psi(r)=\frac{q}{4\pi\epsilon r}\e^{-\kappa_D r}.
\end{equation}

The situation is different when the source charge moves on a time scale comparable to ionic relaxation. The ionic cloud then lags behind the charge and is distorted. This effect is well known in electrophoresis, where polarization and relaxation of the diffuse layer modify the particle mobility \cite{OBrienWhite_1978}. Related distortions occur in sedimentation, shear flow, and constant-velocity translation \cite{Booth_JCP_1954,Lever_1979,Adar_PRE_2018}. In all these cases, the imposed motion selects a direction and the average ionic cloud is anisotropic.

Our question is whether a charged swimmer is screened in the same way as a passive charge. The swimmer has no mean propulsion direction, and its average ionic cloud remains isotropic. Nevertheless, as the cloud responds to the full trajectory, the ballistic-to-diffusive crossover of the swimmer may leave an imprint on the average screening cloud.

\begin{figure}[tpb]
    \centering
    \includegraphics[width=0.75\linewidth]{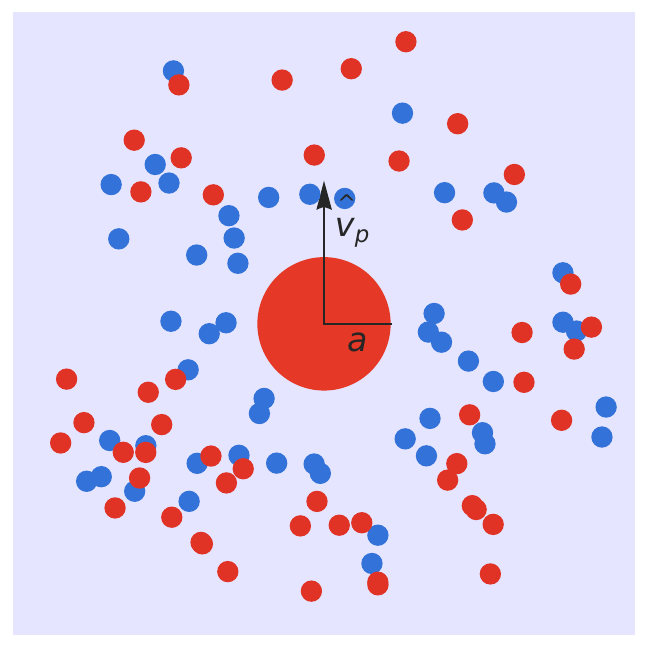}
    \caption{Schematic of a charged active swimmer of radius $a$ in a dilute electrolyte. The arrow denotes the instantaneous propulsion velocity of typical magnitude $v_p$. Cations and anions are shown in red and blue, respectively.}
    \label{fig:model_diagram}
\end{figure}

In this work, we consider a charged swimmer whose self-propulsion velocity follows an Ornstein--Uhlenbeck process \cite{Martin_AOUP_PRE_2021}. Starting from the coupled Nernst--Planck, Poisson, and Stokes equations, we calculate the average steady-state ionic charge distribution around the swimmer. At low activity, we recover the DH profile. At high activity, persistence produces additional spatial regimes. At intermediate distances, the charge density can exhibit a power-law behavior $\sim1/R^2$ followed by a stretched-exponential decay at intermediate distances. For distances exceeding the persistence length, the Yukawa form is recovered with a screening length increased by the swimmer's long-time diffusivity. Thus, activity changes both the form of screening at intermediate distances and its range in the far field. Our results imply that charged swimmers may interact over considerable length scales despite the presence of mobile ions, motivating the study of ``active electrolytes'' composed of positively and negatively charged swimmers.

\section{\label{sec:model}Model}
We consider an electrolyte solution composed of monovalent cations and anions, with respective number densities $n_{+}(\mathbf{r}, t)$ and $n_{-}(\mathbf{r}, t)$. Their average bulk concentration is given by $n_b$. The ions move at velocities $\mathbf{v}_{+}(\mathbf{r}, t)$ and $\mathbf{v}_{-}(\mathbf{r}, t)$ within an incompressible fluid of viscosity $\eta$ with the velocity flow field $\mathbf{u}(\mathbf{r})$. The charge density $c(\mathbf{r}, t)$ and number density $n(\mathbf{r}, t)$ are given by
\begin{align}\label{ion_charge_density}
    c(\mathbf{r}, t)=en_+(\mathbf{r}, t) - en_-(\mathbf{r}, t), \\
    n(\mathbf{r}, t)=n_+(\mathbf{r}, t) + n_-(\mathbf{r}, t),
\end{align}
while the electric current and number-current densities are
\begin{align}\label{current_density}
    \mathbf{j}(\mathbf{r}, t) = e[n_+(\mathbf{r}, t) \mathbf{v}_+(\mathbf{r}, t) - n_-(\mathbf{r}, t) \mathbf{v}_-(\mathbf{r}, t)], \\
    \mathbf{J}(\mathbf{r}, t) = n_+(\mathbf{r}, t) \mathbf{v}_+(\mathbf{r}, t) + n_-(\mathbf{r}, t) \mathbf{v}_-(\mathbf{r}, t).
\end{align}
In the overdamped limit, ions experience a local balance of forces arising from hydrodynamic drag, electrostatic interactions, and pressure gradients, given by:
\begin{equation}\label{ion_balance}
    \xi n_\pm \left( \mathbf{v}_{\pm} - \mathbf{u} \right) = \mp e n_\pm \mathbf{\nabla} \psi - \kbt \mathbf{\nabla} n_\pm,
\end{equation}
where $\xi$ is the ion drag coefficient, assumed to be equal for both cations and anions. The second term on the right-hand side represents the pressure gradient under the ideal-gas approximation, valid for dilute ionic concentrations.

By using Eqs.~\eqref{ion_charge_density} - \eqref{ion_balance}, we readily obtain the ion-charge and number current density,
\begin{align}\label{ion_current_density}
    \mathbf{j} &= c \mathbf{u} - D \mathbf{\nabla} c - \frac{e^2 D}{\kbt} n \mathbf{\nabla} \psi,\\
    \mathbf{J} &= n \mathbf{u} - D \mathbf{\nabla} n - \frac{D }{\kbt} c \mathbf{\nabla} \psi,
\end{align}
where we define the diffusivity of ions $D=k_BT/\xi$. Since total charge and ion numbers are conserved, the spatiotemporal evolution of the local ion charge density $c$ and number density $n$ is governed by the continuity equations:
\begin{align}
\frac{\partial c}{\partial t} &= -\mathbf{\nabla} \cdot \mathbf{j} = D\nabla^2 c - \mathbf{u} \cdot \mathbf{\nabla}c + \frac{e^2 D}{\kbt} \mathbf{\nabla} \cdot \left( n \mathbf{\nabla} \psi \right), \label{ion_charge_continuity} \\
\frac{\partial n}{\partial t} &= -\mathbf{\nabla} \cdot \mathbf{J} = D\nabla^2 n - \mathbf{u} \cdot \mathbf{\nabla}n + \frac{D}{\kbt} \mathbf{\nabla} \cdot \left( c \mathbf{\nabla} \psi \right). \label{ion_number_continuity}
\end{align}
Here, we have used the incompressibility condition $\mathbf{\nabla} \cdot \mathbf{u} = 0$. This coupled system represents the Nernst-Planck equations for charge and number densities of ions in the electrolyte, which depend on the electrostatic potential $\psi$ and the incompressible flow field $\mathbf{u}$ of the solvent. In the presence of external charge density $q$, $\psi$  satisfies Poisson's equation,
\begin{align}\label{poisson_eqn}
\epsilon\nabla^2\psi
=
-q-c.
\end{align}
We consider the fluid to be at a low Reynolds number and incompressible, which leads to the Stokes equation,
\begin{align}
\eta\mathbf{\nabla}^2\mathbf{u} - \mathbf{\nabla} p &= -(\mathbf{f} - c\mathbf{\nabla}\psi), \label{stokes_eq} 
\end{align}
where $p$ is the pressure that enforces incompressibility, and $\mathbf{f}$ and $-c\mathbf{\nabla}\psi$ represent the force densities exerted on the fluid by an external source and electrostatic interactions associated with the ionic charge distribution, respectively.

Eqs.~\eqref{ion_charge_continuity}--\eqref{stokes_eq} constitute the governing equations of the model for an arbitrary source. We now introduce an \emph{active}, spherical, charged particle (swimmer) of radius $a$ and charge $q_0$ as the source in the  (see Fig.~\ref{fig:model_diagram}). We focus on length scales much larger than the particle size $a$ that serves as the smallest length scale in our analysis. For observation length scales $r \gg a$, the external charge density can be written as 
\begin{align}\label{external_charge_density}
    q(\mathbf{r}, t)= q_0 \delta\left[\mathbf{r} - \mathbf{r}_{p}(t)\right],
\end{align}
where $\mathbf{r}_p(t)$ is the position (of the centre of mass) of the charged particle at time $t$. If the charged particle is static, then $\mathbf{r}_p$ is a constant, and there is no source to drive the fluid mechanically, i.e., $\mathbf{f}=0$. This corresponds to the equilibrium limit, and the steady-state solution of Eq.~\eqref{ion_charge_continuity} would yield the classic Debye screening \cite{debye1923theorie1, debye1923theorie2} with the screened electrostatic potential of Eq.~\eqref{debye_huckel_potential}. 

In contrast, we model the position of the active charged particle to evolve according to the following Langevin equation:
\begin{align}\label{rp_dynamics}
\frac{\D \mathbf{r}_p}{\D t} = \mathbf{v}_{a}(t) - \frac{q_{0}}{\xi_p} \mathbf{\nabla}\psi .
\end{align}
Here $\xi_p$ is the translational drag coefficient of the particle. The active velocity $\mathbf{v}_{a}(t)$ is stochastic and is modeled as an Ornstein-Uhlenbeck Process (OUP),
\begin{align}\label{va_dynamics}
\tau_p \frac{\D \mathbf{v}_{a}}{\D t} = - \mathbf{v}_{a} + \sqrt{2D_p} \mathbf{\eta}_{a}(t),
\end{align}
with $\tau_p$ being the particle's characteristic persistence time. $\mathbf{\eta}_{a}(t)$ is a Gaussian white noise vector with zero mean and unit variance, while $D_p$ represents the strength of the active noise. Note that Eq.~\eqref{va_dynamics} allows us to interpret $\mathbf{v}_{a}(t)$ as a colored noise with persistence time $\tau_p$, i.e., it exhibits the following autocorrelation relation:
\begin{align}\label{vel_correlation_AOUP}
\langle \mathbf{v}_{a}(t) \mathbf{v}_{a}(t') \rangle &= v_p^{2} \exp\left(- \frac{|t - t'|}{\tau_p}\right) \mathbf{I}, \\
v_p &= \sqrt{\frac{D_p}{\tau_p}}=\frac{l_p}{\tau_p},
\end{align}
where $v_p$ is the self-propulsion speed of the active particle, and the corresponding persistence length $l_p \equiv  v_p \tau_p$ is then given by,
\begin{align}
    l_p = \sqrt{D_{p} \tau_p}.
    \label{persistence_length}
\end{align}
The different activity parameters can be written in terms of $l_p$ and $\tau_p$ and are summarized in Table 1
\begin{table}[t]
\centering
\begin{tabular}{ccl}
\hline
Quantity & Definition \\
\hline
Persistence length
& $l_p$
 \\[6pt]
Persistence time
& $\tau_p$ \\[6pt]
Long-time diffusion coefficient
& $ D_p = l_p^2/\tau_p$ \\[6pt]
Typical propulsion speed
& $ v_p = l_p/\tau_p$ \\
\hline
\end{tabular}
\caption{Relations among the active particle's persistence length, persistence time,
diffusion coefficient, and velocity. Only two of these quantities are
independent.}
\label{tab:swimmer-parameters}
\end{table} 

The active particle is a swimmer, propelling through the solvent. In the limit of negligible inertia, the propulsive forces balance the opposing hydrodynamic drag exerted by the fluid. Consequently, the swimmer is force-free and torque-free, exerting no net force or torque on the fluid \cite{Batchelor_1970, Lauga_2009, Elgeti_2015,arnoulx2024anomalous, Adhikari_PRL_2014}. In the far field, the leading-order fluid disturbance is generated by an effective force dipole, the stresslet $\mathbf{S}$, and the corresponding force density field $\mathbf{f}$  is given by
\begin{equation}\label{force_density_dipole}
\mathbf{f}(\mathbf{r},t) = -\mathbf{S}\cdot\mathbf{\nabla}\delta\left(\mathbf{r}-\mathbf{r}_p(t)\right).
\end{equation}
The stresslet $\mathbf{S}$ is a rank 2 symmetric and traceless tensor. For swimmers having an axis of symmetry along the instantaneous propulsion direction $\boldsymbol{\widehat{v}_a}$, $\mathbf{S}$ can be written as \cite{Lauga_PRL_2016},
\begin{align}\label{stresslet_squirmer}
    \boldsymbol{S} = a^2 S_{0} \left( \boldsymbol{\widehat{v}_a}\boldsymbol{\widehat{v}_a} - \frac{1}{3}\mathbf{I} \right).
\end{align}
Thus, it is proportional to the square of the size of the swimmer and the value of $S_0$, which depends on the details of the swimming mechanism.

\section{\label{sec:linearization}Linearized equations}
The ion charge and number densities $c$ and $n$, respectively, in Eqs.~\eqref{ion_charge_continuity} and \eqref{ion_number_continuity} are coupled to each other.
In this section, we prescribe a linearization scheme that allows us to decouple them and solve for the ion charge density at steady state. To this end, we consider the source charge $q_{0}$ as a small perturbation and expand local fields $\Phi = \Phi_{0} + \Phi_{1}$ up to linear order in $q_0$. We do not expand in the activity parameters, such as $\tau_p$ and $D_p$. 

As $q_0$ generates the ionic charge density and electrostatic potential, $c_0=0$ and $\psi_0=0$. The zeroth-order number density $n_0$ obeys
\begin{align}
   \frac{\partial n_0}{\partial t} = D\nabla^2 n_0 - \mathbf{u}_0\cdot\mathbf{\nabla}n_{0},
   \label{n0_eq_real_linear}
\end{align} 
which consists of the diffusion of ions and the advection from the solvent flow. Assuming a homogeneous electrolyte at $t=0$, i.e., $n_0(\mathbf{r}, 0)= 2n_b$, we find that $n_0(\mathbf{r}, t)= 2n_b$ at all times.

Next, we turn to the first-order correction to the potential and charge density. The first-order potential $\psi_{1}$ satisfies Poisson's equation,
\begin{align}\label{poisson_eqn_linear}
    \epsilon \nabla^2 \psi_{1}(\mathbf{r}, t ) = - q(\mathbf{r}, t) - c_{1}(\mathbf{r}, t),
\end{align}
and the first-order charge density $c_{1}$ obeys the following equation:
\begin{align}
   \frac{\partial c_1}{\partial t} = D(\nabla^2 - \kappa_D^2) c_1 - \mathbf{u_0}\cdot\mathbf{\nabla}c_{1} - D \kappa_D^2 q.
   \label{c1_eq_real}
\end{align}


The first-order ion charge density $c_{1}$ has an advection term due to the zeroth-order solvent flow $\mathbf{u_0}$, generated by the activity of the swimmer alone. At large length scales, $\mathbf{u_0}$ satisfies the Stokes equation,
\begin{align}
 \eta \nabla^2 \mathbf{u_0} - \nabla p &= -\mathbf{f}, \label{Stokes_zero}
   \end{align}
where $\mathbf{f}$ is the force density which is driven by the activity of the swimmer, as shown in Eq.~\eqref{force_density_dipole}. The resulting flow field is given by (see Appendix~\ref{app:stresslet_derivation} for more details)  $\mathbf{u}_0 \sim a^2/(r - r_p)^2$ for $r\gg a$. Therefore, far away from the swimmer, the advection term $\mathbf{u}_0 \cdot \mathbf{\nabla}$ in Eq.~\eqref{c1_eq_real} is negligible compared to the diffusion term, and we are left with
\begin{align}
   \frac{\partial c_1}{\partial t} \simeq D(\nabla^2 - \kappa_D^2) c_1  - D \kappa_D^2 q.
   \label{c1_eq_real_1}
\end{align}

For simplicity, we non-dimensionalize the equations. We rescale lengths by the Debye  length and times with the characteristic timescale to diffuse across a Debye length (ambipolar time) $\tau_D=\lambda_D^2/D$, as follows:
\begin{align}
r \to \widetilde{r} = r \kappa_D,
\qquad
t \to \widetilde{t} = \frac{t}{\tau_D}.
\end{align}
 The activity parameters are similarly made dimensionless as
\begin{align}
\widetilde{l}_p &= l_p\kappa_D,\\
\widetilde{\tau}_p &= \frac{\tau_p}{\tau_D}.
\end{align}
To simplify the notation, we henceforth drop the tildes from all dimensionless variables and activity parameters. 
In the dimensionless form, Eq.~\eqref{c1_eq_real_1} is given by
\begin{align}
    \frac{\partial c_1}{\partial t} = (\nabla^2 - 1) c_1 - q_0 \delta[\mathbf{r} - \mathbf{r}_{p,0}(t)],
   \label{c1_eq_real_2}
\end{align}
By construction, Eq.~\eqref{c1_eq_real_2} is linear in the swimmer charge $q_0$. The swimmer's activity is manifested in the active dynamics of $r_{p,0}(t)$. It evolves as an OUP with $\psi_0=0$ [Eq.~\eqref{rp_dynamics}]. 

Next, we explore the steady-state solution of the above equation and simplify it in various limits.

\section{Results}
 In Fourier space, $c_{1}$ admits the following solution:
\begin{align}\label{c1_k_labframe}
    \widehat{c}_{1}(\mathbf{k},t) = - q_0 \int_{0}^{t} dt' e^{-(k^2 + 1)(t-t')} e^{-i \mathbf{k}\cdot \mathbf{r}_{p,0}(t')}.
\end{align}
Note that the above solution gives the time-dependent charge-ion concentration, $\widehat{c}_{1}$, in Fourier space as measured in the lab frame. In the reference frame of the swimmer, the corresponding solution, $\widehat{C}_{1}$, can be obtained through the following transformation:
\begin{align}\label{frame_transformation}
    \widehat{C}_{1}(\mathbf{k},t) = \widehat{c}_{1}(\mathbf{k},t) e^{i \mathbf{k}\cdot \mathbf{r}_{p,0}(t)}.
\end{align}
By using Eqs.~\eqref{frame_transformation} and \eqref{c1_k_labframe}, we finally obtain the solution in the swimmer's frame of reference 
\begin{align}
    \widehat{C}_{1}(\mathbf{k},t) = - q_{0} \int_{0}^{t} d\tau e^{-(k^2 + 1)\tau} e^{i \mathbf{k}\cdot \delta \mathbf{r}_{p,0}(\tau)},
    \label{c1_k_chargeframe}
    \end{align}
    where $\delta \mathbf{r}_{p,0}(\tau)= \mathbf{r}_{p,0}(t) -  \mathbf{r}_{p,0}(t'=t-\tau)$ is the displacement of the swimmer. 
    
    Notably, Eq.~\eqref{c1_k_chargeframe} corresponds to a single realization of the active noise. Upon averaging, it takes the following form:
\begin{align}\label{C1_comoving_avrg}
    \langle \widehat{C}_{1}(\mathbf{k},t) \rangle = - q_{0} \int_{0}^{t} d\tau e^{-(k^2 + 1)\tau} \langle e^{i \mathbf{k}\cdot \delta \mathbf{r}_{p,0}(\tau)} \rangle.
\end{align}
As the active velocity $\mathbf{v}_{a}$ follows OU dynamics driven by Gaussian white noise [see Eq.~\eqref{va_dynamics}], the displacement $\delta \mathbf{r}_{p,0}(\tau)$ obeys Gaussian statistics.  Consequently, its characteristic function reduces to
\begin{align}\label{characteristic_function}
    \langle e^{i \mathbf{k}\cdot \delta \mathbf{r}_{p,0}(\tau)} \rangle = e^{-\frac{k^2}{2}\langle \delta r_{p.0}^{2} (\tau) \rangle},
\end{align}
where the variance of displacement along a single Cartesian direction is given by 
\begin{align}\label{msD_poup}
    \langle \delta r_{p.0}^{2} (\tau) \rangle = 2 l_p^2 \left[\frac{\tau}{\tau_p} -  \left( 1 - e^{-\tau/\tau_p} \right) \right]. 
\end{align}
By combining Eqs.~\eqref{characteristic_function} and \eqref{msD_poup} in Eq.~\eqref{C1_comoving_avrg} and taking the limit $t \to \infty$, we finally obtain the average steady-state distribution $\langle \widehat{C}_{1}(\mathbf{k})\rangle_{ss} \equiv \lim_{t \to \infty}\langle \widehat{C}_{1}(\mathbf{k},t) \rangle$ of the ion charge density in the Fourier space, 
\begin{align}\label{C1_ss_rescale}
    \langle \widehat{C}_{1}(\mathbf{k}) \rangle_{ss} &= - q_{0} \int_{0}^{\infty} d\tau e^{-S(k, \tau)}. 
    \end{align}
    Here we have defined the effective action
       \begin{align}\label{actions}
    S(k, \tau) &= \alpha(\tau) k^2 + \tau, \\ 
     \alpha(\tau) &= \tau(1+l_p^2/\tau_p) -  l_p^2 \left( 1 - e^{-\tau/\tau_p} \right). \label{alpha1}
    \end{align}

      We obtain the real-space  charge density in the swimmer's frame of reference by performing the inverse Fourier transform
\begin{align}\label{c1_r}
    \langle C_{1}(\mathbf{R}) \rangle_{ss} &= \frac{1}{(2 \pi)^3}\int d\mathbf{k} e^{i\mathbf{k} \cdot \mathbf{R} } \langle \widehat{C}_{1}(\mathbf{k}) \rangle_{ss}.
\end{align}
Here $\mathbf{R}=\mathbf{r}-\mathbf{r}_{p,0}$ is the position vector relative to the swimmer.
By using Eqs.~\eqref{C1_ss_rescale} and \eqref{c1_r} we calculate $ \langle C_{1}(\mathbf{R}) \rangle_{ss}$ for $d=3$ (see  Appendix~\ref{app:C1_real_solution}) and obtain
\begin{equation}\label{C1_solution_3d}
  \langle C_{1}(\mathbf{R}) \rangle_{\mathrm{ss}} = -\frac{q_{0}}{8 \pi^{3/2}} \int_{0}^{\infty} \frac{\mathrm{d}\tau}{\alpha(\tau)^{3/2}} \exp\left[-\left(\tau + \frac{R^2}{4\alpha(\tau)}\right)\right].  
\end{equation}

Eq.~\eqref{C1_solution_3d} describes the effect of activity on the average steady-state charge density around a charged swimmer, and it constitutes the main result of this work. In order to explicitly solve the integral in several limits, we identify the lag time $\tau_{\star}$ that maximizes the integrand for a given spatial separation $R$. Maximizing the integrand of Eq.~\eqref{C1_solution_3d}, we find that
\begin{align}\label{dominant_lag_condition}
    R^{2}
    =
    6\alpha(\tau_{\star})
    +
    \frac{
        4\alpha^{2}(\tau_{\star})
    }{
        \alpha'(\tau_{\star})
    },
\end{align}
where the prime denotes differentiation with respect to $\tau$.

As the right-hand side of 
Eq.~\eqref{dominant_lag_condition} is a monotonically increasing function, the
observation distance $R$ increases with the dominant lag time $\tau_\ast$. Consequently, the short- and
long-time dynamics of the active particle determine the steady-state
charge-density profile at short and large distances, respectively. Here and throughout, ``short distances" refer to
distances shorter than the relevant dynamical length scales, while remaining much larger than the swimmer radius.

Next we estimate the integral of Eq.~\eqref{C1_solution_3d} in different limits, determined by the separation $R$, activity parameters, and their corresponding dominant lag time. To do so, we approximate $\alpha(\tau)$ of Eq.~\eqref{alpha1} and evaluate the integral of Eq.~\eqref{C1_solution_3d} accordingly. We then determine the spatial range for which this approximation is valid by the self consistency of Eq.~\eqref{dominant_lag_condition}. First, we show that relatively passive swimmers with $D_p\ll1$ induce the standard equilibrium ionic cloud, while active swimmers display a modified ionic charge density.

\subsection{Low-diffusivity regime: $D_p\ll 1$}\label{sec:low_diffusivity}
We first consider the regime where the long-time diffusion of the charged swimmer is smaller than that of the ions. This means that, over long timescales, ions move more rapidly than the source charge. As a result, activity is negligible, and we expect the equilibrium Debye-H\"uckel result to hold. 

To show this, we use the identity $l_p^{2}=D_p\tau_p$ in the exact expression of $\alpha(\tau)$, given in Eq.~\eqref{alpha1}, and find that it is bounded as follows:
\begin{align}
\tau
\leq
\alpha(\tau)
\leq
\left(1+D_p\right)\tau,
\end{align}
for $D_p \geq 0$ and any $\tau$ (Appendix~\ref{app:alpha_small_Da}).  The relative width of the above bound is $D_p$ and therefore becomes small as $D_p$ decreases. Consequently, in the limit $D_p \ll 1$, we obtain, uniformly for all $\tau$,
\begin{align}
\alpha(\tau)\simeq\tau.
\end{align}
Since Eq.~\eqref{dominant_lag_condition} establishes a monotonic correspondence between the dominant lag time $\tau_{\star}$ and the observation distance $R$, the same leading-order approximation $\alpha(\tau)\simeq\tau$ applies at all spatial distances. Substituting this approximation into Eq.~\eqref{C1_solution_3d}, we obtain 
\begin{align}
\left\langle
C_1(R)
\right\rangle_{\mathrm{ss}}
&\simeq
-\frac{q_0}{8\pi^{3/2}}
\int_0^\infty
d\tau
\tau^{-3/2}
\exp\left(
-\tau
-\frac{R^{2}}{4\tau}
\right)
\nonumber\\
&=
-\frac{q_0}{4\pi R}
e^{-R}.
\label{low_activity_DH}
\end{align}
This is the standard Debye--H\"uckel charge-density profile. Thus, in the limit of small $D_p$, the leading-order charge-density profile generated by the active source coincides, at all spatial distances, with that generated by a static point charge. 
Note that the full exponent includes a correction of order $D_p R$ that is negligible compared to the $-R$ term, but may be important for numerical evaluation.

\subsection{High diffusivity regime: $D_p \gg 1$}\label{sec:large_diffusivity}
Here the long-time diffusivity of the active charged source exceeds that of the ions. Consequently, $\alpha(\tau)$ can no longer be described by a single asymptotic form over the entire range of lag times. Instead, it exhibits three distinct behaviors:
\begin{eqnarray}\label{alpha_asymptotic_largeDa}
  \alpha(\tau) \simeq
\left\{
\begin{array}{ll}
\tau, &  \tau \ll 2v_p^{-2},
\vspace{0.25 cm}
\\
\frac{1}{2} v_p^2\tau^2, &  \tau_p \gg \tau \gg 2v_p^{-2},
\vspace{0.25 cm}
\\
 \tau (1 + D_p)  - D_p\tau_p, &   \tau \gg \tau_p.
\end{array}
\right.
\end{eqnarray}
Using $D_p=v_p^{2}\tau_p\gg 1$, it can be shown that the two characteristic time scales in Eq.~\eqref{alpha_asymptotic_largeDa} are well separated, i.e., $\tau_p \gg 2v_p^{-2}$. 

Since the dominant lag time $\tau_{\star}$ increases with the observation distance $R$,  these three dynamical regimes correspond, respectively, to short, intermediate, and large observation distances. We show below that these regimes produce an equilibrium-like Yukawa profile of the charge density, an intermediate stretched-exponential profile with an algebraic decay as a limiting form, and an underscreened Yukawa profile, respectively.

\subsubsection{Short-distance regime}\label{sec:large_D_short_scale}
At short distances, the dominant lag time satisfies $\tau_{\star}\ll2v_p^{-2}$, for which
$\alpha(\tau)
\simeq
\tau.$
The corresponding calculation is therefore identical to that leading to Eq.~\eqref{low_activity_DH}, and the charge-density profile is given by the equilibrium DH form,
\begin{align}\label{DH_short}
\left\langle
C_1(R)
\right\rangle_{\mathrm{ss}}
\simeq
-\frac{q_0}{4\pi R}
e^{-R}.
\end{align}
Thus, even in the large-diffusivity regime, the charge-density profile remains equilibrium-like at sufficiently short distances.

To obtain the spatial range for which the above charge density profile holds, we use $\alpha(\tau_{\star})= \tau_{\star}$ in Eq.~\eqref{dominant_lag_condition} and enforce the condition $\tau_{\star} \ll 2v_p^{-2}$. This yields the spatial range of the equilibrium-like profile
\begin{eqnarray}\label{DH_short_range}
  R \ll
\left\{
\begin{array}{ll}
v_p^{-1}, &  v_p \gg 1,
\vspace{0.25 cm}
\\
v_p^{-2}, &  v_p \ll 1.
\end{array}
\right.
\end{eqnarray} 
The short-distance regime thus depends on the typical velocity of the swimmer $v_p$. For high velocities, $R\ll1$ and thus no screening is observed in this regime, but rather the bare Coulomb interaction. For low velocities, however, $v_p^{-2}\gg1,$ and screening may be observed.

\subsubsection{Intermediate-distance regime}\label{sec:large_D_mid_scale}
The intermediate regime is dominated by the characteristic lag time $\tau_{\star}$ satisfying $\tau_p \gg \tau_{\star} \gg 2v_p^{-2}$. The corresponding $\alpha(\tau)$ is then dominated by the ballistic mode and is given by \begin{align}
    \alpha(\tau)
    \simeq
    \frac{v_p^{\,2}\tau^{\,2}}{2}.
    \label{ballistic_alpha}
\end{align}
Substituting Eq.~\eqref{ballistic_alpha} in Eq.~\eqref{C1_solution_3d}, we find the charge density profile
\begin{align}
    \langle
        C_1(R)
    \rangle_{\mathrm{ss}}
    =
    -
    \frac{q_0}
    {2^{3/2}\pi^2 v_p R^{\,2}}
    \,
G_{0,3}^{3,0}\!\left(
\frac{R^{2}}{8v_p^2}
\,\middle|\,
0,\frac{1}{2},1
\right),
    \label{ballistic_scaling_form}
\end{align}
given in terms of the Meijer $G$-function, defined as
\begin{eqnarray}
    \,
G_{0,3}^{3,0}\!\left(
\frac{x^{2}}{8}
\,\middle|\,
0,\frac{1}{2},1
\right)
    =
    \sqrt{\pi}\int_{0}^{\infty}
    dy\,y^{-3}
    \exp\left(
        -xy-\frac{1}{2y^{2}}
    \right).
    \label{ballistic_scaling_function}
\end{eqnarray}
Next, we consider the two asymptotic limits $x\gg1$ and $x\ll1$ for $x=R/v_p$.

\paragraph{ $R/v_p \gg 1:$}
In this case, the Meijer $G$ function shows the following asymptotic behavior:
\begin{equation}
 \lim_{x \to \infty}   \,
G_{0,3}^{3,0}\!\left(
\frac{x^{2}}{8}
\,\middle|\,
0,\frac{1}{2},1
\right)
    \simeq \pi\sqrt{\frac{2}{3}}
x^{1/3}
\exp\left[
-\frac{3}{2}
x^{2/3}
\right],
\label{G-fn_x_large}
\end{equation}
which leads to the charge-density profile
\begin{align}
    \langle
        C_1(R)
    \rangle_{\mathrm{ss}}
    \simeq
    -
    \frac{q_0}
    {2\sqrt{3}\pi
    v_p^{4/3}R^{5/3}} \exp{\left[-\frac{3}{2} \left(\frac{R}{v_p}\right)^{2/3} \right]}.
    \label{stretched_expo_charge_profile}
\end{align}
This is a new regime where the charge density decays in a stretched-exponential fashion. This regime is valid for (Appendix~\ref{app:spatial_extent}) 
\begin{align}\label{spatial_full_ballistic}
    \max\left(
        v_p,
        v_p^{-2}
    \right)
   & \ll
    R
    \ll
    v_p\tau_p^{3/2},\\
    \tau_p & \gg 1.
\end{align}

\paragraph{$R/v_p \ll 1:$}
The Meijer $G$ function approaches $\sqrt{\pi}$ in the limit $x\ll1$, which corresponds to $R \ll v_p$. In the steady state, the corresponding charge density profile is given by
\begin{align}
    \langle
        C_1(R)
    \rangle_{\mathrm{ss}}
    \simeq
    -
    \frac{q_0}
    {(2\pi)^{3/2}
    v_pR^{\,2}}.
    \label{scale_free_charge_profile}
\end{align}
Therefore, in this regime, the steady-state charge density decays algebraically as $1/R^{2}$, and the magnitude scales inversely with $v_p$. This reflects that, as the swimmer's speed increases, ions have less time to respond, leading to reduced ionic rearrangement in the vicinity of the source.
We now insert Eq.~\eqref{ballistic_alpha} into Eq.~\eqref{dominant_lag_condition}, along with the condition $\tau_p \gg \tau_{\star} \gg 2v_p^{-2}$, to obtain the spatial validity of the power-law behavior obtained in Eq.~\eqref{scale_free_charge_profile}. We find that, in addition to $D_p\gg1$, to conditions should be met (Appendix~\ref{app:spatial_extent}):
\begin{align}
v_p^{-1}
& \ll
R
\ll
\min\left(
l_p,v_p
\right),
\label{ballistic_spatial_window} \\
v_p & \gg1.
\label{large_vs}
\end{align}
%
\begin{figure}
    \centering
    \includegraphics[width=1\linewidth]{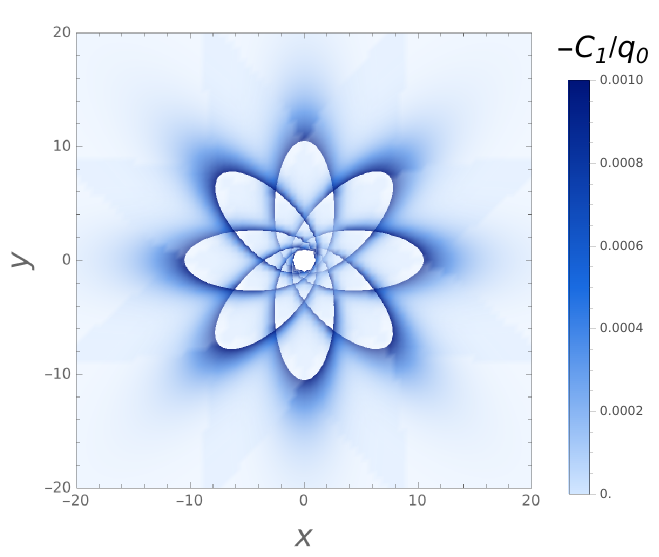}
    \caption{Two-dimensional projection of the charge density profile across eight different active propulsion directions. For each direction, the profile is strongly anisotropic and exhibits a comet-like structure as given by Eq.~\eqref{C_angle_anisotropy}. The isotropic density profile in Eq.~\eqref{scale_free_charge_profile} is obtained by averaging over all such orientation realizations.}
    \label{fig:scale_free_c1}
\end{figure}

To understand the $1/R^2$ decay in Eq.~\eqref{scale_free_charge_profile}, we present an alternative heuristic argument. At length scales smaller than the persistence length ($R \ll l_p$), the charged swimmer undergoes persistent, effectively ballistic motion along a single propulsion direction. For a specific noise realization, the swimmer moves uniformly with an approximate velcotiy of magnitude $v_p$ along a fixed direction. This maps the local dynamics onto the problem of a charge moving at constant velocity through an electrolyte, as studied in Ref.~\cite{Adar_PRE_2018}. In that case, the uniform velocity breaks spherical symmetry and generates an anisotropic, comet-shaped ionic cloud. In the reference frame of the moving source, the steady-state charge profile at position $\mathbf{R}$ is given by:
\begin{equation}\label{C_angle_anisotropy}
C(\mathbf{R}, \theta) = - \frac{q_0}{4\pi R} \exp\left[ \frac{v_p R \cos\theta}{2} - \sqrt{1 + \frac{v_{p}^2}{4}} R \right],
\end{equation}
where $\theta$ is the angle between the velocity and $\mathbf{R}$. Because the swimmer has no preferred direction, the isotropic ensemble average involves integrating over a uniform angular distribution: 
\begin{align}
C(R) &= \frac{1}{2} \int_0^\pi \sin\theta\,C(\mathbf{R}, \theta)  d\theta \nonumber \\
&= - \frac{q_0}{2\pi v_p R^2} \sinh\left( \frac{v_p R}{2} \right) \exp\left(-\sqrt{1 + \frac{v_p^2}{4}} R\right).
\end{align}
An illustration of this averaging procedure is given in Fig.~\ref{fig:scale_free_c1}. In the limit of high velocities ($v_{p} \gg 1$), $\sqrt{1 + v_{p}^2/4} \approx v_{p}/2$, leading to
\begin{equation}
C(R) \sim - \frac{q_0}{v_p R^2}.
\end{equation}
This recovers the  $\sim1/(v_p R^2)$ charge density of Eq.~\eqref{scale_free_charge_profile}.

\subsubsection{Long-distance regime}\label{sec:large_D_large_scale}
In the long-distance regime, we have the dominant lag time $\tau_{\star}$ much larger than the persistence time $\tau_p$. For $\tau \gg \tau_p$, we approximate
\begin{eqnarray}\label{alpha_underscreening}
    \alpha(\tau) \simeq \tau \lambda^2 - D_p \tau_p, 
\end{eqnarray}
where $\lambda^2 = 1 + D_p$. We obtain the charge density profile by substituting $\alpha(\tau)$ in Eq.~\eqref{C1_solution_3d}. The resulting long-distance profile is given by (Appendix~\ref{app:long_distance_profile})
\begin{align}
    \langle
        C_1(R)
    \rangle_{\mathrm{ss}}
    \simeq
    -
    \frac{q_0}
    {4\pi\lambda^{2}R}
    e^{-\tau_p}\exp\left(
        -\frac{R}{\lambda}
    \right).
    \label{underscreened_yukawa}
\end{align}
Thus, the long-distance steady-state profile has the Yukawa form with a prefactor that is exponentially suppressed with increasing persistence time $\tau_p$.

We now determine the spatial region over which this profile is valid. To this end, we note that the lag time $\tau_\star \gg \tau_p$ and as a result we approximate $\alpha(\tau_\star) \simeq \lambda^2 \tau_{\star}$. Using these in the dominance criterion in Eq.~\eqref{dominant_lag_condition}, the spatial regime is calculated to be,
\begin{eqnarray}\label{underscreening_range}
  R \gtrsim
\left\{
\begin{array}{ll}
\lambda \sqrt{\tau_p}, &  \tau_p \ll 1,
\vspace{0.25 cm}
\\
\lambda \tau_p, &  \tau_p \gg 1.
\end{array}
\right.
\end{eqnarray}

In the small-persistence limit, the exponential prefactor in Eq.~\eqref{underscreened_yukawa} approaches unity, and the steady-state profile reduces to a Yukawa form with the effective screening length $\lambda = \sqrt{1+D_p}$. Since $D_p\gg1$, we have $\lambda\gg1$, implying that the screening length is enhanced relative to the equilibrium Debye length. The electrolyte is therefore \textit{underscreened}. Moreover, in this small-persistence regime, we can write the onset of the underscreened regime  $\lambda \sqrt{\tau_p}\simeq \sqrt{D_p\tau_p}=l_p$. Hence, the underscreened profile begins at distances of the order of the persistence length.

However, in the large persistence limit for which we have $\tau_p \gg 1$, we find that the amplitude of the Yukawa profile, which begins at a length scale larger than $\lambda \tau_p \simeq l_p \sqrt{\tau}_p$, is exponentially vanishing with $\tau_p$. Thus, activity-induced underscreening is physically relevant when the persistence time is comparable to or shorter than the ionic relaxation time. For much longer persistence times, this effect becomes difficult to observe.

\subsection{Regime diagram}
\label{sec:phase_diagram}

\begin{figure*}[t]
    \centering
    \subfloat[
    \label{fig:phase_diagram_small_tau}]
    {%
        \includegraphics[width=0.49\linewidth]
        {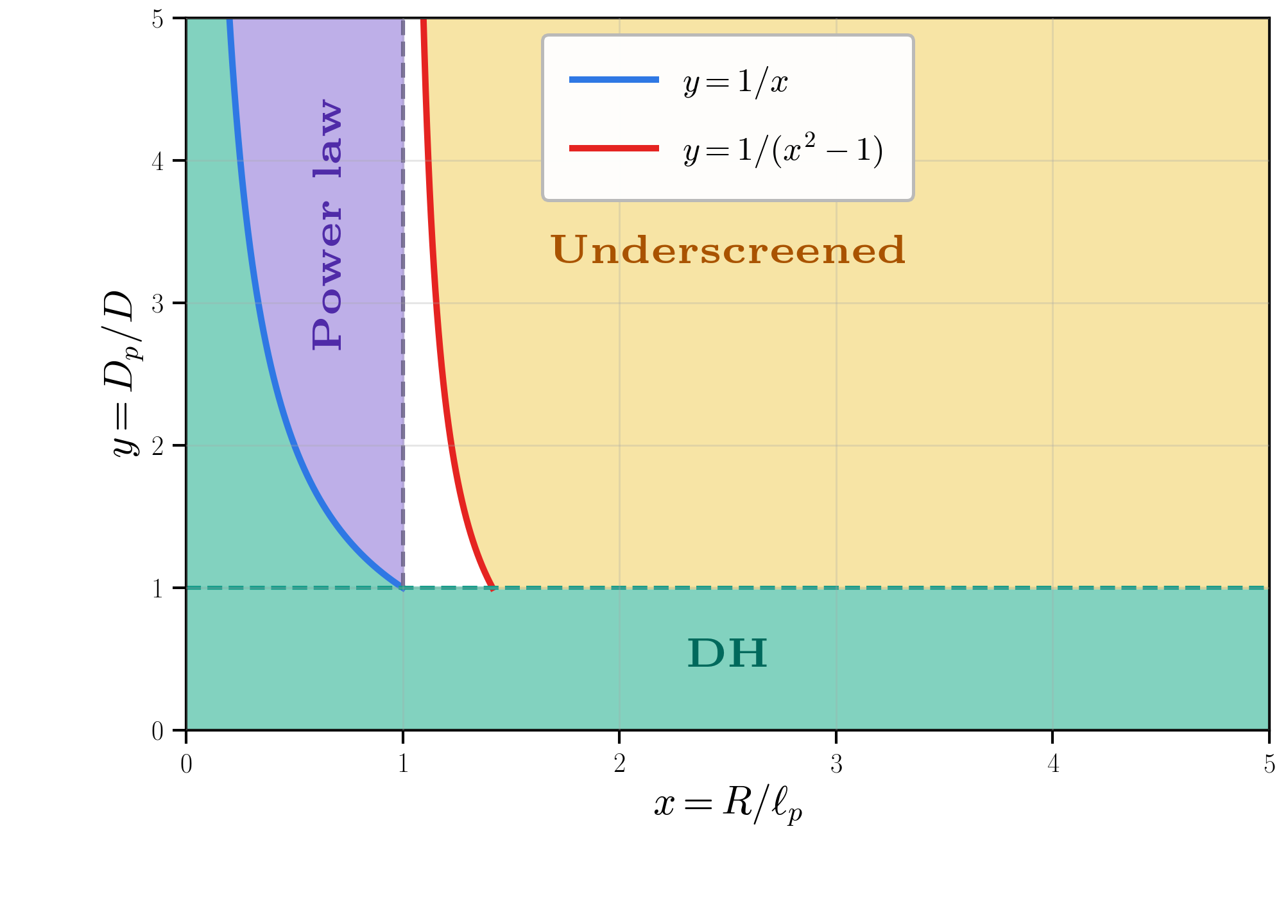}
    }
    \hfill
    \subfloat[
    \label{fig:phase_diagram_large_tau}]
    {%
        \includegraphics[width=0.49\linewidth]
        {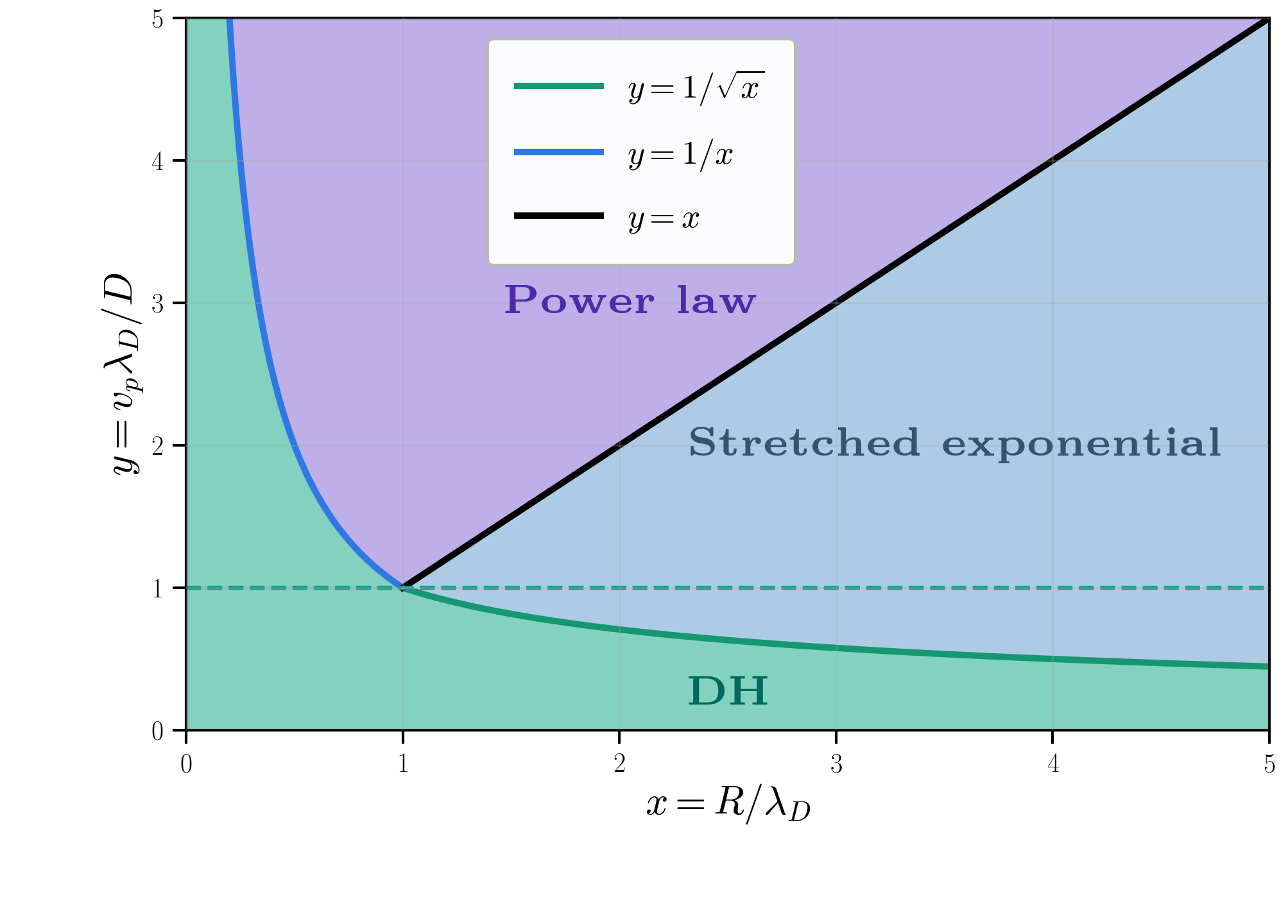}
    }
    \caption{
        Regime diagrams of the charge-density profile. Panel (a) shows the different regimes in the $R/l_p$-$D_p/D$ plane for small persistence time, while panel (b) shows the corresponding regimes in the $R/\lambda_D$-$v_p\lambda_D/D$ plane for large persistence time. 
    }
    \label{fig:phase_diagrams}
\end{figure*}
Next, we summarize the distinct possible spatial behaviors of the charge-density profile. We consider two limiting regimes of the (dimensionless) persistence time $\tau_p$ and identify the corresponding regions in an activity-distance plane. 

\subsubsection{Short persistence time: $\tau_p \lesssim 1$}
In this case, the swimmer dynamics mostly appears to be diffusive and the relevant activity parameter is the diffusivity $D_p$, which we denote by $y$. The spatial scale is given by $x=R/l_p$. The short-persistence regime diagram in the $(x,y)$ plane is depicted in Fig.~\ref{fig:phase_diagram_small_tau}. In the figure, the dimensionless coordinates are written explicitly as \(R/l_p\) and \(D_p/D\), with all quantities expressed in dimensional form.  

As shown in Sec.~\ref{sec:low_diffusivity}, for
$D_p\ll1$, the charge-density profile retains the equilibrium
Debye--H\"uckel (DH) form at all observation distances. This result is also independent of $\tau_p$. We therefore identify
the entire low-activity region $y<1$ as the DH
regime, shown in green in Fig.~\ref{fig:phase_diagram_small_tau}. Here,
$y=1$ is used as a schematic crossover boundary.

For $y=D_p \gg 1$ and small $\tau_p$, the
dimensionless self-propulsion speed satisfies
$v_p\gg1$. As shown in Sec.~\ref{sec:large_diffusivity}, the charge density in this case exhibits three distinct behaviors at successively larger distances: an inner Debye--H\"uckel
screening regime [see Eq.~\eqref{DH_short}], an intermediate power-law regime [see Eq.~\eqref{scale_free_charge_profile}], and an asymptotic
underscreened regime [see Eq.~\eqref{underscreened_yukawa}].

Using Eq.~\eqref{DH_short_range}, the spatial range of the inner
Debye--H\"uckel regime can be expressed in terms of $x$ and $y$ as
\begin{align}
1\ll y\ll\frac{1}{x}.
\end{align}
The curve $y=1/x$ is plotted in blue in Fig.~\ref{fig:phase_diagram_small_tau} and the region below it is shaded accordingly in green. 
We now use Eq.~\eqref{ballistic_spatial_window} and find the spatial range of the power-law behavior. In the limit of $\tau_p \lesssim 1$, it satisfies the following relation:
\begin{align}
\frac{1}{y}\ll x\ll 1,
\end{align}
with $y\gg 1$. This region lies above the blue crossover curve $y=1/x$ and to the
left of $x=1$, and is shaded in purple in
Fig.~\ref{fig:phase_diagram_small_tau}.

Finally, using Eq.~\eqref{underscreening_range}, we obtain the spatial range for the underscreened profile,
\begin{align}
y\gg \frac{1}{x^2 -1}.
\end{align}
This region is shaded in yellow, above the red crossover curve $y=1/(x^2-1)$ in Fig.~\ref{fig:phase_diagram_small_tau}. The blank portion of the diagram represents an intermediate
 region that is not characterized by the limiting cases considered here.

\subsubsection{Long-persistence time: $\tau_p\gg1$}
In the long-persistence regime, the swimmer mostly appears to move ballistically and the natural activity parameter is the active velocity $v_p$. Accordingly, the regime diagram is represented in Fig.~\ref{fig:phase_diagram_large_tau} using $x=R$ and $y=v_p$ for $\tau_p\gg1$. In the figure, the dimensionless coordinates are written explicitly as \(R/\lambda_D\) and \(v_p \lambda_D/D\), with all quantities expressed in dimensional form. We focus on the joint limit $D_p\gg1$ and $\tau_p\gg1$, with
$
v_p=\sqrt{D_p/\tau_p}
$
remaining finite. The small $D_p \ll 1$ limit yields the conventional Debye-H\"uckel profile at all distances. At fixed $\tau_p$, this limit corresponds to $v_p \to 0$ and is therefore contained in the small-$v_p$ sector of the regime diagram.

The analysis in Sec.~\ref{sec:large_D_short_scale} shows that the
charge-density profile exhibits Debye--H\"uckel screening at short
distances: $R \ll v_p^{-2}$ for
$v_p\ll1$ and
$R \ll v_p^{-1}$ for
$v_p\gg1$ [see Eq.~\eqref{DH_short_range}]. In the
$x$--$y$ plane, these conditions correspond to the region below the
blue curve $y=1/x$ for $y>1$ and below the green curve
$y=1/\sqrt{x}$ for $y<1$. We shade the resulting
Debye--H\"uckel region in green in
Fig.~\ref{fig:phase_diagram_large_tau}.

For the intermediate-scale power-law behavior discussed in
Sec.~\ref{sec:large_D_mid_scale}, the corresponding spatial window
is given by Eq.~\eqref{ballistic_spatial_window}. In the
large-$\tau_p$ limit, this can be equivalently expressed as
\begin{align}
    y^{-1}\ll x\ll y,
    \qquad y\gg1.
\end{align}
We shade the corresponding region, lying above both the blue curve
$y=1/x$ and the black curve $y=x$, in purple. 

In the intermediate range, we also have the stretched exponential regime given in Eq.~\eqref{stretched_expo_charge_profile}. The spatial range corresponding to this regime follows Eq.~\eqref{spatial_full_ballistic}, which can be rewritten as,
\begin{align}
\max{(y, y^{-2})}\ll x\ll y\tau_p^{3/2}.
\end{align}
In the formal large-persistence limit
$\tau_p\to\infty$, the upper spatial bound
$y\tau_p^{3/2}$ diverges and therefore does not define
a finite boundary in the regime diagram. The remaining condition can
then be expressed as
\begin{align}
    x^{-1/2}\ll y\ll x.
\end{align}
Thus, this region lies above the green curve $y=1/\sqrt{x}$ and
below the black curve $y=x$ and is shaded in blue. We refer to this region as the stretched exponential regime. 

Finally, in the formal large-persistence limit $\tau_p\to\infty$, the onset distance of the underscreened regime [Eq.~\eqref{underscreening_range}] diverges, and also the amplitude of the profile vanishes exponentially. The underscreened regime, therefore, does not appear in the regime diagram.

\section{Summary and concluding remarks}
In this paper, we have developed a linearized framework to characterize the average steady-state ionic charge density around a charged swimmer in a dilute electrolyte. Compared with the equilibrium DH result, where charges are screened within a Debye length $\lambda_D$, activity can lead to new regimes, depending on the interplay between three dimensionless variables (written here explicitly with their proper dimensions): the distance from the swimmer $R/\lambda_D$, the persistence length $l_p/\lambda_D$, and the persistence time $\tau_p/\tau_D$.
Two 2D cuts of this 3D space are presented in Fig.~\ref{fig:phase_diagrams}


In the low-activity regime, the source moves slowly compared with the ionic relaxation timescale. The ions effectively experience the swimmer as quasistatic, and classical DH screening is recovered. At higher activity, the interplay between ionic relaxation and the persistent motion of the swimmer produces several distinct spatial behaviors. Simply put, the swimmer can either ``outrun'' the ions (high persistence regime, $\tau_p\gg1$) or "shake off" the ions (low persistence regime, $\tau_p\ll1$). Explicitly, in addition to the short-distance DH regime, the intermediate-distance regime can exhibit a power-law decay followed by a stretched-exponential decay as the distance from the swimmer increases. Both behaviors arise for large dimensionless diffusivity of the swimmer, with the power-law and stretched-exponential forms additionally requiring large dimensionless self-propulsion speed and large dimensionless persistence time, respectively.  At sufficiently large distances, the profile again becomes exponential with a screening length $\lambda/\lambda_D=\sqrt{1+D_p}$, indicating activity-induced underscreening. For large persistence, however, this underscreened profile is strongly suppressed, and its onset is shifted to increasingly large distances, making the regime practically unobservable.

Closely related phenomena arise in phoretic active matter~\cite{golestanian2022phoretic,IllienGolestanian_2024}. Finite solute diffusion makes the propulsion of a chemically active particle depend on its recent orientational history, reflecting the delayed rearrangement of its self-generated solute cloud. At the collective level, coarse-grained theories of interacting phoretic particles lead to a generalized Poisson--Boltzmann description with an inverse screening length
$\lambda\sim\sqrt{D_{\mathrm{eff}}})$, formally resembling our far-field result. The physical origin is, however, different. In the phoretic problem the screening medium is itself composed of active particles, whereas here a single active charge is screened by passive electrolyte ions. Moreover, the long-time collective description of the phoretic system does not capture the persistence-controlled stretched-exponential and algebraic regimes found here. 


An interesting question is whether the regimes described above can be observed experimentally. To address this, we estimate the relevant parameters. Conventional self-propelled particles typically exhibit persistence times in the range \(0.1<\tau_p<5\,\mathrm{s}\), while their self-propulsion speeds and active diffusivities are estimated to be of order \(1\text{--}10\,\mathrm{\mu m\,s^{-1}}\) and \(0.1\text{--}100\,\mathrm{\mu m^2\,s^{-1}}\), respectively \cite{BERG1972,Howse_PRL_2007, Jiang_PRL_2010}. The corresponding ionic scales are the diffusivity $D$, the characteristic diffusive velocity $D/\lambda_D$, and the ambipolar relaxation time $\tau_D=\lambda_D^2/D$. In aqueous electrolytes at room temperature, $D\sim10^3\,\mathrm{\mu m^2\,s^{-1}}$ \cite{vanysek1993ionic}.  An upper bound on $\tau_D$ and lower bound on the  velocity $D/\lambda_D$ are obtained for the large possible $\lambda_D\simeq1\mathrm{\mu m}$, obtained at infinite dilution at pH $7$. Consequently, $\tau_D\sim10^{-3}\,\mathrm{s}$ and $D/\lambda_D\sim10^3\,\mathrm{\mu m\,s^{-1}}$. Thus, both the ionic diffusivity and the characteristic ionic velocities exceed their active-particle counterparts, while the persistence time $\tau_p$ greatly exceeds the ambipolar time $\tau_D$. These estimates suggest that, aside from the classic DH regime, only the stretched-exponential one can be observed for sufficiently large distances. However, for $v_p$ of the order $10\mathrm{\mu m\,s^{-1}}$ this distance is estimated as $R\simeq1\mathrm{cm}$. Alternatively, to observe the effect at a distance of the order $R\simeq1\,\mathrm{\mu m}$, the active velocity should be $10^3\mathrm{\mu m\,s^{-1}}$.
We conclude that only the classical DH regime is observable under typical aqueous conditions with currently available swimmer velocities.

Rather than increasing the swimmer velocity, it is possible to lower the charge carriers' diffusivity and concentration, when they are not simple ions. This arises in nonpolar electrolytes, where charged reverse micelles act as mobile charge carriers and have been reported to produce much larger Debye lengths, reaching $\lambda_D\simeq10\,\mathrm{\mu m}$ \cite{hsu2005charge,Roberts_2008}. Using the measured micellar sizes and solvent viscosities, we estimate their diffusivities to be of the order of $100\,\mathrm{\mu m^2\,s^{-1}}$. Near the upper end of the reported Debye-length range, the characteristic speed and ambipolar time are estimated as $D/\lambda_D\simeq10\,\mathrm{\mu m\,s^{-1}}$ and $\tau_D\simeq1\,\mathrm{s}$, respectively. Therefore, it is possible for the active diffusivity and self-propulsion speed to exceed the ionic diffusivity and characteristic ionic speed. Moreover, the persistence time can be either shorter or longer than the ambipolar time. Taken together, these estimates suggest that the intermediate power-law and stretched-exponential regimes, as well as the underscreening regime at small persistence times, may be experimentally accessible. This would require realizing swimmers in this environment, where the activity is not driven by electric fields.

In conclusion, this work establishes a new mechanism through which active motion modifies electrostatic screening. Particularly striking is that, at high propulsion speeds, activity can evade conventional screening by mobile ions and produce a power-law regime. An important direction for future work is to investigate the consequences of this regime in dilute suspensions of charged active particles. The associated incomplete ionic screening may mediate effective long-range interactions between particles whose direct interactions would otherwise be negligible, potentially giving rise to new forms of collective organization. In aqueous experimental environments, however, only the Debye--H\"uckel regime is accessible. 
Nevertheless, the \emph{collective} behavior of the active system is expected to be richer because of the nontrivial response of active particles to potentials~\cite{kafri2026active}.
\\
{\it Acknowledgments.} This article is dedicated to the memory of Prof. Rudolf (Rudi) Podgornik, whose seminal contributions left a lasting mark on soft-matter physics. Beyond his scientific achievements, Rudi was an exceptional mentor and collaborator who profoundly influenced many young scientists, including two of us (TM and RMA). He played an important role in TM’s scientific development during his PhD, while RMA had the privilege of working with him as a doctoral student. Rudi'’s boundless curiosity, remarkable physical intuition, creativity, and enthusiasm made working with him both inspiring and deeply rewarding. We remain deeply grateful for his lasting influence on how we think about and approach science. He will be greatly missed, both as a scientist and as a person.

\appendix
\section{Zeroth-order flow velocity}
\label{app:stresslet_derivation}
The zeroth-order incompressible solvent flow field $\mathbf{u}_0$ satisfies the Stokes equation,
\begin{align}
    \eta \nabla^2 u_{0,\alpha}(\mathbf{r}) - \partial_\alpha p(\mathbf{r}) &= -f_\alpha(\mathbf{r}), \label{eq:stokes_mom} \\
    \partial_\alpha u_{0,\alpha}(\mathbf{r}) &= 0. \label{eq:stokes_incomp}
\end{align}
The force $f_\alpha$ is generated by the activity of the swimmer. Far away from the swimmer, it is given by
\begin{align}\label{f_density}
    f_{\alpha}(\mathbf{r},t) = -S_{\alpha \beta} \partial_{\beta}\delta\left(\mathbf{r}-\mathbf{r}_p(t)\right),
\end{align}
where $S_{\alpha\beta}$ is the stresslest defined in Eq.~\eqref{stresslet_squirmer}. Here $\mathbf{r}$ and $\mathbf{r}_p(t)$ represent the position vector of the point of observation and that of the swimmer from the lab frame. We define the instantaneous relative position vector from the swimmer's center of mass $\mathbf{R}=\mathbf{r}-\mathbf{r}_p(t)$.

In Fourier space, the Stokes equation with the force density field from Eq.~\eqref{f_density} reads as
\begin{align}
    -\eta k^2 \widehat{u}_{0, \alpha}(\mathbf{k}) - i k_\alpha \widehat{p}(\mathbf{k}) &= i S_{\alpha \beta} k_\beta, \label{eq:k_mom_moving} \\
    i k_\alpha \widehat{u}_{0, \alpha}(\mathbf{k}) &= 0. \label{eq:k_incomp_moving}
\end{align}
Applying the transverse projector operator $P_{\alpha \beta}(k)=\delta_{\alpha \beta} - k_\alpha k_\beta/k^2$ eliminates the pressure term and we get,
\begin{equation}
    \widehat{u}_{0, \alpha}(\mathbf{k}) = -\frac{i S_{\alpha \beta} k_\beta}{\eta k^2} + \frac{i S_{\beta \gamma} k_\alpha k_\beta k_\gamma}{\eta k^4}.
\end{equation}

We now perform the inverse Fourier transform to get back the solution in real space. Using the standard 3D identities $\mathcal{F}^{-1}\{k^{-2}\} = (4\pi R)^{-1}$ and $\mathcal{F}^{-1}\{k^{-4}\} = -R / (8\pi)$, we obtain,
\begin{equation}
    u_{0, \alpha}(\mathbf{R}) = -\frac{S_{\alpha \beta}}{\eta} \frac{\partial}{\partial R_\beta} \left( \frac{1}{4\pi R} \right) + \frac{S_{\beta \gamma}}{8\pi \eta} \frac{\partial^3 R}{\partial R_\alpha \partial R_\beta \partial R_\gamma}.
\end{equation}
The first term is given by,
\begin{equation}
    -\frac{S_{\alpha \beta}}{\eta} \frac{\partial}{\partial R_\beta} \left( \frac{1}{4\pi R} \right) = \frac{S_{\alpha \beta} R_\beta}{4\pi \mu R^3},
\end{equation}
and the second term is given by,
\begin{align}
    & \frac{S_{\beta \gamma}}{8\pi \eta} \frac{\partial^3 R}{\partial R_\alpha \partial R_\beta \partial R_\gamma} = \frac{S_{\beta \gamma}}{8\pi \eta} \Bigg(\frac{3 R_\alpha R_\beta R_\gamma}{R^5}   -\frac{\delta_{\beta \gamma} R_\alpha + \delta_{\alpha \beta} R_\gamma + \delta_{\alpha \gamma} R_\beta}{R^3} \Bigg).
\end{align}
We now use the fact that the stresslet tensor is a symmetric and traceless rank 2 tensor. This eventually yields,
\begin{equation}
     \frac{S_{\beta \gamma}}{8\pi \eta} \frac{\partial^3 R}{\partial R_\alpha \partial R_\beta \partial R_\gamma} = -\frac{S_{\alpha \beta} R_\beta}{4\pi \mu R^3} + \frac{3 S_{\beta \gamma} R_\alpha R_\beta R_\gamma}{8\pi \eta R^5}.
\end{equation}
Adding both terms, we get
\begin{equation}
    u_{0, \alpha}(\mathbf{R}) = \frac{3 S_{\beta \gamma} R_\alpha R_\beta R_\gamma}{8\pi \eta R^5}.
\end{equation}
Now using the fact that stresslet scales as square of the swimmer radius, i.e., $S_{\beta \gamma} \sim a^2$, we essentially show that the far field flow velocity decays as $a^2/R^2$.

\section{Real space solution in $3d$}\label{app:C1_real_solution}
In this section, we present the derivation of the real-space solution $\langle \widehat{C}_{1}(\mathbf{r})\rangle_{ss}$ in the steady state. By using Eqs.~\eqref{C1_ss_rescale} and \eqref{c1_r}, we obtain 
\begin{align}
     \langle C_{1}(\mathbf{R}) \rangle_{ss} &= -\frac{q_0}{(2 \pi)^3}\int d\mathbf{k} e^{i \mathbf{k} \cdot \mathbf{R}} \int_{0}^{\infty} d\tau e^{-S(k, \tau)}, \nonumber \\
     &= \frac{i q_0}{(2\pi)^2 R} \int_{0}^{\infty} d\tau \int_{0}^{\infty} dk k (e^{i k R} - e^{-i k R}) e^{-S(k, \tau)}, \nonumber \\
     &= \frac{i q_0}{(2\pi)^2R} \int_{0}^{\infty} d\tau \left[I(\tau, R) - I(\tau, -R)\right],
     \label{C1_3d}
\end{align}
where $I(\tau, R)$ is defined as,
\begin{align}
    I(\tau, R) &= \int_{0}^{\infty} dk k e^{i k R} e^{-S(k, \tau)}, \nonumber \\
    &= e^{-\tau}\frac{\partial}{\partial (i R)} \int_{0}^{\infty} dk e^{i k R -\alpha k^2}, \nonumber \\
     &= \frac{\sqrt{\pi}e^{-\tau}}{2\sqrt{\alpha}} \frac{\partial}{\partial (i R)} \left[e^{-\frac{R^2}{4 \alpha}} \left\{1 + erf\left( \frac{iR}{2 \sqrt{\alpha}} \right) \right\} \right], \nonumber \\
      &= \frac{\sqrt{\pi}e^{-\tau}}{2 i\sqrt{\alpha}} \left[\frac{i}{\sqrt{\pi \alpha}} - \frac{R e^{-\frac{R^2}{4 \alpha}}}{2 \alpha} \left(1 +   erf\left( \frac{iR}{2 \sqrt{\alpha}} \right)\right) \right]. \nonumber \\
      \label{I_r}
    \end{align}
  By using $I(\tau,R)$ from Eq.~\eqref{I_r} in Eq.~\eqref{C1_3d}, we obtain
    \begin{align}
         \langle C_{1}(\mathbf{R}) \rangle_{ss} &= -\frac{q_0}{8 \pi^{3/2}} \int_{0}^{\infty} \frac{\D\tau}{\alpha^{3/2}}  e^{-\left(\tau+\frac{R^2}{4 \alpha}\right)}, 
    \end{align}
    where we have defined 
    \begin{align}
        \alpha(\tau) &= \tau(1+D_p) -  l_p^2 \left( 1 - e^{-\tau/\tau_p} \right).
    \end{align}

\section{$\alpha(\tau)$ in the small $D_p$ limit}\label{app:alpha_small_Da}    
We rewrite $\alpha$ as
\begin{align}
\alpha(\tau)
=
\tau
+
D_p
\left[
\tau
-
\tau_p
\left(
1-e^{-\tau/\tau_p}
\right)
\right].
\label{alpha_low_activity_app}
\end{align}
The expression inside the square brackets admits the integral representation
\begin{align}
\tau
-
\tau_p
\left(
1-e^{-\tau/\tau_p}
\right)
=\int_0^{\tau}
ds
\left(
1-e^{-s/\tau_p}
\right).
\end{align}
Since $0\leq 1-e^{-s/\tau_p}\leq 1$ for $s\geq0$, it follows that
\begin{align}
0
\leq
\tau
-
\tau_p
\left(
1-e^{-\tau/\tau_p}
\right)
\leq
\tau.
\end{align}
Consequently,
\begin{align}
\tau
\leq
\alpha(\tau)
\leq
\left(1+D_p\right)\tau.
\end{align}
Notably, the bound derived above is valid for all $D_p \geq 0$. Its relative width is $D_p$, and therefore becomes small as $D_p$ decreases. Consequently, in the limit $D_p \ll 1$, we obtain, uniformly for all $\tau$,
\begin{align}
\alpha(\tau)\simeq\tau.
\end{align}

\section{Spatial extent of the intermediate regime}\label{app:spatial_extent}
To determine the spatial range over which the intermediate ballistic regime is observed, we substitute
\begin{equation}
\alpha(\tau) = \frac{1}{2} v_{p}^2 \tau^2
\end{equation}
into the dominance criterion,
\begin{equation}
R^{2} = 6\alpha(\tau_{\star}) + \frac{4\alpha^{2}(\tau_{\star})}{\alpha'(\tau_{\star})}.
\end{equation}
This gives
\begin{align}
    \tau_{\star}^{\,2}
    \left(
        \tau_{\star}+3
    \right)
    =
    \left(
        \frac{R}
        {v_p}
    \right)^2.
    \label{ballistic_saddle}
\end{align}
We now solve Eq.~\eqref{ballistic_saddle} in the limiting cases of small and large $R/v_p$. 

\paragraph{Small $R/v_p$:}
We obtain the $1/R^2$ decay in the limit
$R/v_p\ll1$,
one then has $\tau_{\star}\ll1$, and hence
\begin{align}
    \tau_{\star}
    \sim
    \frac{R}
    {v_p}.
\end{align}
The condition
$\tau_{\star}\gg v_p^{-2} $,
required for the quadratic term in $\alpha(\tau)$ to dominate, gives the lower spatial bound
\begin{align}
    R
    \gg
    v_p^{-1}.
\end{align}
Similarly, the persistence condition
$\tau_{\star}\ll\tau_p$
gives
\begin{align}
    R
    \ll
    v_p\tau_p
    =
    l_p.
\end{align}
Combining these conditions with
$R/v_p\ll1$, we obtain
\begin{align}\label{appendix_power_law_window}
    v_p^{-1}
    \ll
    R
    \ll
    \min\left(
        l_p,v_p
    \right).
\end{align}
For the power-law regime in Eq.~\eqref{appendix_power_law_window} to be physically accessible, its lower bound must lie below both upper bounds. The first requirement is
\begin{align}
v_p^{-1}
\ll
l_p.
\end{align}
Using
$l_p=v_p\tau_p$ and
$D_p=v_p^{\,2}\tau_p$,
this condition becomes
\begin{align}
v_pl_p
=
D_p
\gg1,
\end{align}
which is already guaranteed in the large-diffusivity regime
considered here. The second requirement is
\begin{align}
v_p^{-1}
\ll
v_p,
\end{align}
or equivalently,
\begin{align}
v_p^{\,2}\gg1.
\end{align}
Therefore, a parametrically well-defined power-law window requires
\begin{align}
D_p\gg1,
\qquad
v_p\gg1.
\end{align}
For $v_p>1$, a finite interval can in principle exist,
whereas a clear separation between its lower and upper bounds
requires $v_p\gg1$. By contrast, when $v_p\ll1$, the two necessary conditions $R \gg v_p^{-1}$ and $R \ll v_p$
are mutually incompatible, and the power-law result is not
self-consistent in this limit.

\paragraph{Large $R/v_p$:} Using Eq.~\eqref{ballistic_saddle} in this case yields,
\begin{align}
    \tau_{\star} \sim \left(\frac{R}{v_p}\right)^{2/3}.
\end{align}
The condition $\tau_{\star}\gg v_p^{-2}$,
required for the quadratic term in $\alpha(\tau)$ to dominate, yields the lower spatial bound
\begin{align}
    R
    \gg
    v_p^{-2}.
\end{align}
Also, as we are in the regime $R \gg v_p$, so the lower bound of $R$ is then given by,
\begin{align}
     R
    \gg \max{(
    v_p^{-2}, v_p)}.
\end{align}
In order to calculate the upper bound, we use the fact $\tau_\star \ll \tau_p$, which gives
\begin{align}
    R
    \ll
    v_p\tau_p^{3/2}.
\end{align}
So, the spatial window in this case is given by,
\begin{align}
    \max{(
    v_p^{-2}, v_p)} \ll R
    \ll
    v_p\tau_p^{3/2}.
\end{align}
For this spatial window to be consistent, both lower bounds must be smaller than the upper bound. First,
\begin{align}
 v_p
\ll
 v_p\tau_p^{3/2},
\end{align}
which suggests
\begin{align}
    \tau_p\gg 1.
\end{align}
Similarly,
\begin{align}
 v_p^{-2}
&\ll
 v_p\tau_p^{3/2}, \nonumber
\\
\left( v_p^2\tau_p\right)^{3/2}
&=
 D_p^{3/2}
\gg1.
\end{align}
Thus, the spatial window is self-consistent, provided 
\begin{align}
\tau_p\gg1,
\qquad
 D_p\gg1.
\end{align}
This is exactly the large-persistence and large-diffusivity limit considered here.

\section{Derivation of the underscreened profile}
\label{app:long_distance_profile}

We begin with the integral expression for the steady-state charge-density profile, as shown in the main text in Eq.~\eqref{C1_solution_3d},
\begin{align}
\left\langle C_1(R)\right\rangle_{\mathrm{ss}}
=
-\frac{q_0}{8\pi^{3/2}}
\int_0^\infty
\frac{d\tau}{\alpha(\tau)^{3/2}}
\exp\left[
-\tau-\frac{R^2}{4\alpha(\tau)}
\right].
\label{density_integral_appendix}
\end{align}
In the long-distance regime, the integrand is dominated by $\tau_\star \gg \tau_p$. As a result, the leading-order contribution to the above integral is obtained from the region $\tau \gg \tau_p$; we therefore neglect the contribution from $\tau \ll \tau_p$, and obtain
\begin{align}
\left\langle C_1(R)\right\rangle_{\mathrm{ss}}
\simeq -\frac{q_0}{8\pi^{3/2}}
\int_{\tau_{p}}^{\infty}
\frac{d\tau}{\alpha(\tau)^{3/2}}
\exp\left[
-\tau-\frac{R^2}{4\alpha(\tau)}
\right].
\label{density_integral_appendix_1}
\end{align}
In the long-time regime, i.e., for $\tau\gg\tau_p$, we use the approximated form 
\begin{align}
\alpha(\tau)
\simeq
\lambda^2\tau-l_p^2
=
\lambda^2\left(\tau-\tau_0\right),
\qquad
\tau_0\equiv\frac{l_p^2}{\lambda^2},
\label{alpha_long_time_appendix}
\end{align}
where $\lambda^2=1+D_p$. Introducing
the shifted variable,
\begin{align}
s=\tau-\tau_0,
\end{align}
we obtain
\begin{align}
\left\langle C_1(R)\right\rangle_{\mathrm{ss}}
\simeq
-\frac{q_0e^{-\tau_0}}{8\pi^{3/2}\lambda^3}
\int_{s_{\min}}^\infty
ds\,s^{-3/2}
\exp\left[
-s-\frac{R^2}{4\lambda^2s}
\right],
\label{density_shifted_integral}
\end{align}
where
\begin{align}
s_{\min}
=
\tau_p-\tau_0
=
\frac{\tau_p}{\lambda^2}.
\end{align}
In this regime, the dominant lag time satisfies
$\tau_\star\gg\tau_p$, such that the shifted dominant time is much larger
than the lower cutoff, i.e., $s_{\star} \gg s_{\min}$. We may, therefore, extend the lower integration limit
in Eq.~\eqref{density_shifted_integral} to zero. Moreover, in the large $D_p$ limit, we have $\tau_0 \simeq \tau_p$. This gives
\begin{align}
\left\langle C_1(R)\right\rangle_{\mathrm{ss}}
\simeq
-\frac{q_0e^{-\tau_p}}{8\pi^{3/2}\lambda^3}
\int_0^\infty
ds\,s^{-3/2}
\exp\left[
-s-\frac{R^2}{4\lambda^2s}
\right].
\end{align}
Using the identity
\begin{align}
\int_0^\infty
ds\,s^{-3/2}
\exp\left(
-s-\frac{a^2}{4s}
\right)
=
\frac{2\sqrt{\pi}}{a}e^{-a},
\qquad a>0,
\end{align}
with $a=R/\lambda$, we finally obtain
\begin{align}
\left\langle C_1(R)\right\rangle_{\mathrm{ss}}
\simeq
-\frac{q_0}{4\pi\lambda^2R}
e^{-\tau_p}
\exp\left(
-\frac{R}{\lambda}
\right).
\label{underscreened_profile_appendix}
\end{align}
Thus, the long-distance charge-density profile has a Yukawa form with the
enhanced screening length $\lambda=\sqrt{1+D_p}$ and an amplitude suppressed
by the factor $e^{-\tau_p}$.

\bibliographystyle{unsrt}
\bibliography{aipsamp}

\begin{thebibliography}{10}

\bibitem{Ramaswamy_2010}
Sriram Ramaswamy.
\newblock The mechanics and statistics of active matter.
\newblock {\em Annual Review of Condensed Matter Physics}, 1(Volume 1,
  2010):323--345, 2010.

\bibitem{Marchetti_RMP_2013}
M.~C. Marchetti, J.~F. Joanny, S.~Ramaswamy, T.~B. Liverpool, J.~Prost, Madan
  Rao, and R.~Aditi Simha.
\newblock Hydrodynamics of soft active matter.
\newblock {\em Rev. Mod. Phys.}, 85:1143--1189, Jul 2013.

\bibitem{Bechinger_RMP_2016}
Clemens Bechinger, Roberto Di~Leonardo, Hartmut L\"owen, Charles Reichhardt,
  Giorgio Volpe, and Giovanni Volpe.
\newblock Active particles in complex and crowded environments.
\newblock {\em Rev. Mod. Phys.}, 88:045006, Nov 2016.

\bibitem{MIPS_Cates_2015}
Michael~E. Cates and Julien Tailleur.
\newblock Motility-induced phase separation.
\newblock {\em Annual Review of Condensed Matter Physics}, 6(Volume 6,
  2015):219--244, 2015.

\bibitem{Reichhardt_ratchet_2017}
C.J.~Olson Reichhardt and C.~Reichhardt.
\newblock Ratchet effects in active matter systems.
\newblock {\em Annual Review of Condensed Matter Physics}, 8(Volume 8,
  2017):51--75, 2017.

\bibitem{metzger2026revisiting}
Jessica Metzger, Sunghan Ro, and Julien Tailleur.
\newblock Revisiting the ratchet principle: When hidden conservation laws
  prevent directed currents in stochastic systems.
\newblock {\em Physical Review Letters}, 136(25):257102, 2026.

\bibitem{metzger2026revisitingE}
Jessica Metzger, Sunghan Ro, and Julien Tailleur.
\newblock Revisiting the ratchet principle: Lessons from active and passive
  stochastic dynamics.
\newblock {\em Physical Review E}, 113(6):064152, 2026.

\bibitem{Granek_RMP_2024}
Omer Granek, Yariv Kafri, Mehran Kardar, Sunghan Ro, Julien Tailleur, and
  Alexandre Solon.
\newblock Colloquium: Inclusions, boundaries, and disorder in scalar active
  matter.
\newblock {\em Rev. Mod. Phys.}, 96:031003, Sep 2024.

\bibitem{Lauga_2009}
Eric Lauga and Thomas~R Powers.
\newblock The hydrodynamics of swimming microorganisms.
\newblock {\em Reports on Progress in Physics}, 72(9):096601, aug 2009.

\bibitem{Elgeti_2015}
J~Elgeti, R~G Winkler, and G~Gompper.
\newblock Physics of microswimmers—single particle motion and collective
  behavior: a review.
\newblock {\em Reports on Progress in Physics}, 78(5):056601, apr 2015.

\bibitem{TailleurCates_PRL_2008}
J.~Tailleur and M.~E. Cates.
\newblock Statistical mechanics of interacting run-and-tumble bacteria.
\newblock {\em Phys. Rev. Lett.}, 100:218103, May 2008.

\bibitem{Lefranc_PRX_2025}
Thibault Lefranc, Alberto Dinelli, Carla Fern{\'a}ndez-Rico, Roel P.~A.
  Dullens, Julien Tailleur, and Denis Bartolo.
\newblock Synthetic quorum sensing and absorbing phase transitions in colloidal
  active matter.
\newblock {\em Phys. Rev. X}, 15:031050, Aug 2025.

\bibitem{IllienGolestanianSen_CSR_2017}
Pierre Illien, Ramin Golestanian, and Ayusman Sen.
\newblock {'Fuelled'} motion: Phoretic motility and collective behaviour of
  active colloids.
\newblock {\em Chemical Society Reviews}, 46(18):5508--5518, 2017.

\bibitem{IllienGolestanian_2024}
Pierre Illien and Ramin Golestanian.
\newblock Chemotactic particles as strong electrolytes: {D}ebye--{H}{\"u}ckel
  approximation and effective mobility law.
\newblock {\em arXiv preprint arXiv:2402.09775}, 2024.

\bibitem{NasouriGolestanian_PRL_2020}
Babak Nasouri and Ramin Golestanian.
\newblock Exact phoretic interaction of two chemically active particles.
\newblock {\em Phys. Rev. Lett.}, 124:168003, Apr 2020.

\bibitem{AsmolovVinogradova_POF_2023}
Evgeny~S. Asmolov and Olga~I. Vinogradova.
\newblock Limiting propulsion of ionic microswimmers.
\newblock {\em Physics of Fluids}, 35(7):072007, 2023.

\bibitem{ShresthaOlvera_PRE_2024}
Ahis Shrestha and Monica Olvera de~la Cruz.
\newblock Enhanced phoretic self-propulsion of active colloids through surface
  charge asymmetry.
\newblock {\em Phys. Rev. E}, 109(1):014613, Jan 2024.

\bibitem{frydel2018}
Derek Frydel and Rudolf Podgornik.
\newblock Mean-field theory of active electrolytes: Dynamic adsorption and
  overscreening.
\newblock {\em Physical Review E}, 97(5):052609, 2018.

\bibitem{verwey1948theory}
E.~J. Verwey and J.~Th.~G. Overbeek.
\newblock {\em Theory of the Stability of Lyophobic Colloids}.
\newblock Elsevier, New York, 1948.

\bibitem{andelman1995electrostatic}
David Andelman.
\newblock Electrostatic properties of membranes.
\newblock In Reinhard Lipowsky and Erich Sackmann, editors, {\em Handbook of
  Physics of Biological Systems}, volume~1, chapter~12. Elsevier Science,
  Amsterdam, 1995.

\bibitem{levin2002electrostatic}
Yan Levin.
\newblock Electrostatic correlations: from plasma to biology.
\newblock {\em Reports on Progress in Physics}, 65(11):1577--1632, 2002.

\bibitem{safran2003statistical}
Samuel Safran.
\newblock {\em Statistical Thermodynamics of Surfaces, Interfaces, and
  Membranes}.
\newblock CRC Press, Boca Raton, FL, 2003.

\bibitem{israelachvili2011intermolecular}
Jacob~N. Israelachvili.
\newblock {\em Intermolecular and Surface Forces}.
\newblock Academic Press, New York, 3rd edition, 2011.

\bibitem{markovich2021charged}
Tomer Markovich, David Andelman, and Rudolf Podgornik.
\newblock Charged membranes: {P}oisson--{B}oltzmann theory, the {DLVO}
  paradigm, and beyond.
\newblock In Cyrus~R. Safinya and Joachim~O. R{\"a}dler, editors, {\em Handbook
  of Lipid Membranes: Molecular, Functional, and Materials Aspects}, chapter~6.
  CRC Press, Boca Raton, FL, 2021.
\newblock Cited as "in press" in original 2019 text.

\bibitem{debye1923theorie1}
Peter Debye.
\newblock Zur {T}heorie der {E}lektrolyte. {I}. {G}efrierpunktserniedrigung und
  verwandte {E}rscheinungen.
\newblock {\em Physikalische Zeitschrift}, 24:185--206, 1923.

\bibitem{debye1923theorie2}
Peter Debye and Erich H{\"u}ckel.
\newblock Zur {T}heorie der {E}lektrolyte. {II}. {D}as {G}renzgesetz f{\"u}r
  die elektrische {L}eitf{\"a}higkeit.
\newblock {\em Physikalische Zeitschrift}, 24:305--325, 1923.

\bibitem{OBrienWhite_1978}
R.~W. O'Brien and L.~R. White.
\newblock Electrophoretic mobility of a spherical colloidal particle.
\newblock {\em Journal of the Chemical Society, Faraday Transactions 2:
  Molecular and Chemical Physics}, 74:1607--1626, 1978.

\bibitem{Booth_JCP_1954}
F.~Booth.
\newblock Sedimentation potential and velocity of solid spherical particles.
\newblock {\em The Journal of Chemical Physics}, 22(12):1956--1968, 12 1954.

\bibitem{Lever_1979}
D.~A. Lever.
\newblock Large distortion of the electric double layer around a charged
  particle by a shear flow.
\newblock {\em Journal of Fluid Mechanics}, 92(3):421–433, 1979.

\bibitem{Adar_PRE_2018}
Ram~M. Adar, Yuki Uematsu, Shigeyuki Komura, and David Andelman.
\newblock Linear response functions of an electrolyte solution in a uniform
  flow.
\newblock {\em Phys. Rev. E}, 98:032604, Sep 2018.

\bibitem{Martin_AOUP_PRE_2021}
David Martin, J\'er\'emy O'Byrne, Michael~E. Cates, \'Etienne Fodor, Cesare
  Nardini, Julien Tailleur, and Fr\'ed\'eric van Wijland.
\newblock Statistical mechanics of active ornstein-uhlenbeck particles.
\newblock {\em Phys. Rev. E}, 103:032607, Mar 2021.

\bibitem{Batchelor_1970}
G.~K. Batchelor.
\newblock The stress system in a suspension of force-free particles.
\newblock {\em Journal of Fluid Mechanics}, 41(3):545–570, 1970.

\bibitem{arnoulx2024anomalous}
Thibaut Arnoulx~de Pirey, Yariv Kafri, and Sriram Ramaswamy.
\newblock Anomalous long-ranged influence of an inclusion in
  momentum-conserving active fluids.
\newblock {\em Physical Review X}, 14(4):041034, 2024.

\bibitem{Adhikari_PRL_2014}
Somdeb Ghose and R.~Adhikari.
\newblock Irreducible representations of oscillatory and swirling flows in
  active soft matter.
\newblock {\em Phys. Rev. Lett.}, 112:118102, Mar 2014.

\bibitem{Lauga_PRL_2016}
Eric Lauga and S\'ebastien Michelin.
\newblock Stresslets induced by active swimmers.
\newblock {\em Phys. Rev. Lett.}, 117:148001, Sep 2016.

\bibitem{golestanian2022phoretic}
Ramin Golestanian.
\newblock Phoretic active matter.
\newblock {\em Active matter and nonequilibrium statistical physics}, pages
  230--293, 2022.

\bibitem{BERG1972}
HOWARD~C. BERG and DOUGLAS~A. BROWN.
\newblock Chemotaxis in escherichia coli analysed by three-dimensional
  tracking.
\newblock {\em Nature}, 239(5374):500--504, Oct 1972.

\bibitem{Howse_PRL_2007}
Jonathan~R. Howse, Richard A.~L. Jones, Anthony~J. Ryan, Tim Gough, Reza
  Vafabakhsh, and Ramin Golestanian.
\newblock Self-motile colloidal particles: From directed propulsion to random
  walk.
\newblock {\em Phys. Rev. Lett.}, 99:048102, Jul 2007.

\bibitem{Jiang_PRL_2010}
Hong-Ren Jiang, Natsuhiko Yoshinaga, and Masaki Sano.
\newblock Active motion of a janus particle by self-thermophoresis in a
  defocused laser beam.
\newblock {\em Phys. Rev. Lett.}, 105:268302, Dec 2010.

\bibitem{vanysek1993ionic}
Petr Vanysek.
\newblock Ionic conductivity and diffusion at infinite dilution.
\newblock {\em CRC hand book of chemistry and physics}, pages 5--92, 1993.

\bibitem{hsu2005charge}
Ming~F Hsu, Eric~R Dufresne, and David~A Weitz.
\newblock Charge stabilization in nonpolar solvents.
\newblock {\em Langmuir}, 21(11):4881--4887, 2005.

\bibitem{Roberts_2008}
G.~Seth Roberts, Rodrigo Sanchez, Roger Kemp, Tiffany Wood, and Paul Bartlett.
\newblock Electrostatic charging of nonpolar colloids by reverse micelles.
\newblock {\em Langmuir}, 24(13):6530--6541, 07 2008.

\bibitem{kafri2026active}
Yariv Kafri and Julien Tailleur.
\newblock How active field theories couple to external potentials.
\newblock {\em arXiv preprint arXiv:2603.22424}, 2026.

\end{thebibliography}

\end{document}